# Study Of The Electronic Structure In Oxides Using Absorption And Resonant X-Ray Scattering

Yves Joly[1], Elena Nazarenko[1], Emilio Lorenzo[1], Sergio Di Matteo[2], Calogero Rino Natoli[2]

[1]*Laboratoire de Cristallographie, BP166, 38042 Grenoble Cedex 09, France*
[2]*Laboratori Nazionali di Frascati INFN, via E. Fermi 40, C.P. 13, I-00044 Frascati (Roma) Italy*

**Abstract.** Playing with the polarization of the incident (and outgoing) photon beams, the magnetic field and the orientation of the sample, absorption techniques can be very selective probes of the electronic and geometric structures in material. Among the absorption techniques, Resonant X-ray diffraction (RXD) is a spectroscopy where both the power of site selective diffraction and the power of local absorption spectroscopy regarding atomic species are combined to the best. By virtue of the dependence on the core level state energy and the three dimensional electronic structure of the intermediate state, this technique is specially suited to study charge, orbital or spin orderings and associated geometrical distortions. In the case of charge ordering, we exploit the fact that atoms with closely related site symmetries but with barely different charges exhibit resonances at slightly different energies. Here we show that the sensitivity of this effect allows for quantitative estimations of the charge disproportion. Opposite to fluorescence or absorption measurements, the power of diffraction relies on the capability of detecting differences that are even smaller than the inverse lifetime of the core hole level. To account for the uncertainty of the crystallographic structure and the fact that the charge ordering must be disentangled from the associated atomic displacements, a complete methodology is proposed. It needs a very important set of experimental data, first principle simulations and the use of objective confidence factors for comparing experiment and theory. As main example we apply the method to the evidence and quantitative evaluation of the charge ordering in the low temperature phase of magnetite. We found in this case $\pm$ 0.12 and $\pm$ 0.10 charge disproportions between the four iron octahedral sites. Relative sensitivity on spin ordering, Jahn-Teller distortion and orbital ordering is also shown and compared on different transition metal oxide compounds.

## INTRODUCTION

Most X-ray spectroscopy techniques are related to the transition of a core electron up to some empty states above the Fermi level in crystal or above the highest occupied molecular orbital when using the quantum chemistry vocabulary. The process implies the absorption of a photon. When the absorption is real one gets the well know x-ray absorption fine structures (XAFS). When it is resonant, that is when the absorption process is virtual because occurring during the time given by the Heisenberg's uncertainty relation, the measurement is often performed in the diffraction mode. In this case, the so-called resonant x-ray scattering (RXS) (or diffraction anomalous fine structure DAFS) results from the absorption followed by the immediate elastic emission of another photon. The amplitude of both phenomena, real and virtual absorptions, is related to the probability of transition between the core level and the states in the conduction band. Thus they are closely connected and from the point of view of the simulation when one is able to handle the real absorption case, one also, quite easily manages the virtual case. The probability of transition depends first on the initial state, say a core level well known, second, on transition operators depending on the polarization of the light, and third, the most important, on the unoccupied density of state of the material. Thus, it is clear that these techniques are highly sensitive to the electronic structure of the investigated compound. A number of papers have reported on this subject. RXS is a more recent technique than XAFS. Nevertheless for more than ten years, the capacity of this tool to investigate the electronic properties of a wide class of compounds has been shown. The rapid improvement of the capacities of the last generation of synchrotron has been so fast that the experimental side was during a time in advance in front of the theoretical understanding and simulation. This has led to a number of false demonstrations of expected electronic properties as for instance orbital and charge ordering in oxides. Nevertheless RXS is now coming to an adult age. The power of the technique is better understood and calibrated. Cooperative phenomena start to be disentangle with the help of simulations and, more important, the technique seems to be able not only to show the evidence of such or such electronic expected property but also to solve quantitatively the corresponding electronic parameters such as the charge disproportion in charge ordered compound as we are going to show in the following.

After a fast description of the physics governing the absorption process we will give the mathematical tools and the methodology allowing the interpretation of the RXS spectra. Then we shall focus our attention on three examples in metal oxides: the magnetic ordering in the low temperature phase in $V_2O_3$, the polar ferrimagnet $GaFeO_3$ analysis and the quantitative resolution of the charge ordering problem in the low temperature phase of magnetite.

## THEORY AND METHODOLGY

We first give the basic equations involved in the RXS spectroscopy. A special attention is given on the tensor approach to describe the anisotropy of the scattering factors also available in x ray absorption near edge structure (XANES). Then the methodology necessary to disentangle correlated parameters is shown.

### Basis Of The RXS Spectroscopy

In RXS, the global process of photon absorption, virtual photoelectron excitation and photon re-emission, is coherent throughout the crystal, thus giving rise to the usual Bragg diffraction condition:

$$F = \sum_j e^{i\vec{Q}\cdot\vec{R}_j}\left(f_{0j} + f_{mj} + f_j' + if''_j\right) \quad (1)$$

Here $\vec{R}_j$ stands for the position of the scattering ion $j$, $\vec{Q}$ is the diffraction vector and $f_{0j}$ and $f_{mj}$ are the Thomson and magnetic non resonant factors of the $j^{th}$-atom. $f_{mj}$ is estimated following Blum and Gibbs[1] even if the condition of high energy is not fulfilled. $f_j'+if_j''$ is the resonant, or anomalous, atomic scattering factor. It is this term highly dependant on energy around the absorption edge and anisotropic in most case that make makes the strength of the technique. At resonance, the photon energy $\hbar\omega$ is closed to $E_g$ - $E_n$, the difference between the core and intermediary state energies. In this condition, with a good approximation, the scattering factor is given by the expression:

$$f_j' - if_j'' = m_e\omega^2 \sum_n \frac{\langle\psi_g|O_o^*|\psi_n\rangle\langle\psi_n|O_i|\psi_g\rangle}{\hbar\omega - (E_n - E_g) + i\frac{\Gamma_n}{2}} \quad (2)$$

where $m_e$ is the electron mass, $\psi_g$ and $\psi_n$ are the core and intermediary state wave functions, and $\Gamma_n$ is the broadening resulting from the life time of the core and intermediary states. The sum is extended over all the excited states of the system. The operator $O$ is written through the multipolar expansion of the photon field up to electric dipole (E1) and quadrupole (E2) terms:

$$O_{i(o)} = \vec{\varepsilon}_{i(o)} \cdot \vec{r}\left(1 + \tfrac{1}{2}i\vec{k}_{i(o)} \cdot \vec{r}\right) \quad (3)$$

where $\vec{r}$ is the electron position and $\vec{\varepsilon}_{i(o)}$ and $\vec{k}_{i(o)}$ are the polarization and wave vector of the incoming (i) and outgoing (o) photons. When incoming and outgoing photon has the same polarization, the imaginary part $f_j''$ is proportional to the absorption cross section.

### Tensor Approach

It is highly convenient to expand the matrix product in the numerator of equation (2), separating the dipole-dipole, dipole-quadrupole and quadrupole-quadrupole factors. Then for this numerator and expanding the polarization and wave vectors in cartesian coordinates, one gets:

$$\begin{aligned}N = &\sum_{\alpha\alpha'}\varepsilon_\alpha^{s*}\varepsilon_{\alpha'}^e D_{\alpha\alpha'} \\ &+ i\sum_{\alpha\alpha'\alpha''}\varepsilon_\alpha^{s*}\varepsilon_{\alpha'}^e\left(k_{\alpha''}^e I_{\alpha\alpha'\alpha''} - k_{\alpha''}^s I_{\alpha'\alpha\alpha''}^*\right) \\ &+ \sum_{\alpha\alpha'\alpha''\alpha'''}\varepsilon_\alpha^{s*}\varepsilon_{\alpha'}^e k_{\alpha''}^s k_{\alpha'''}^e Q_{\alpha\alpha''\alpha'\alpha'''}\end{aligned}$$
(4)

where respectively $D$, $I$ and $Q$ are the E1-E1, E1-E2 and E2-E2 cartesian tensors. The indices $\alpha$ represent x, y, z the Cartesian directions. An equivalent approach uses the spherical tensors. It is less intuitive but more direct for an interpretation in physical terms. In this way one gets:

$$N = \sum_{\substack{\ell=0,2 \\ m=-\ell,\ell}} (-1)^{\ell+m} T_\ell^m D_\ell^m$$

$$+ i \sum_{\substack{\ell=1,3 \\ m=-\ell,\ell}} (-1)^{\ell+m} T_\ell^m I_\ell^m + \sum_{\substack{\ell=0,4 \\ m=-\ell,\ell}} (-1)^{\ell+m} T_\ell^m Q_\ell^m$$

(5)

Here the $T$ are the spherical tensors of the polarization and wave vectors. In the Table 1, we give the names and transformation of the spherical tensors under symmetry inversion and time reversal:

**TABLE 1.** Name and characteristic of the different spherical tensors. The cells of the non existing tensors are ached. The magnetic ones are in yellow. In the cells, the signs ++, -+, +- or – show respectively the changes under, first, time reversal and second, inversion. For some of the tensors the corresponding physical property is given.

|              | ℓ | E1-E1     |    | E1-E2            |    | E2-E2              |    |
|--------------|---|-----------|----|------------------|----|--------------------|----|
| Monopole     | 0 | Charge    | ++ |                  |    | Charge             | ++ |
| Dipole       | 1 | Moment    | -+ |                  | +- | Toroidal moment -- | Related to moment -+ |
| Quadrupole   | 2 |           | ++ | Toroidal axis    | +- | --                 |    | ++ |
| Octupole     | 3 |           |    |                  | +- | --                 |    | -+ |
| Hexadecaople | 4 |           |    |                  |    |                    |    | ++ |

The main interest of this presentation is that it shows that playing with the polarization of the incoming and outgoing photon, inverting the magnetic field, considering the symmetry of the material relying the different absorbing atoms, choosing with care the reflections in order to have Bragg factor operating new selections, one can perform measurements sensitive only to some specific tensor components, that is to very specific projection of the density of state and thus probe precise electromagnetic properties of the material. The well know example is the circular dicroism at the $L_{23}$ edge which select the magnetic dipole (E1-E1) component giving thus through the sum rules a measurement of the magnetic moment. Other examples given in the following will concern the measurement of the toroidal moment.

## Simulation

The calculations are performed according to state of the art codes, i.e, using the fully relativistic code (FDMNES)[2,3] that has already been efficient in the simulation of different spin, orbital, charge or geometrical ordering phenomena[4,5]. It uses the multiple scattering theory or alternatively the finite difference method to calculate the anomalous structure factors. It is also specially adapted for the RXS purpose in order to get the different reflections in a user friendly way and gives an analysis in term of spherical tensors.

## Methodology

When parameters must be solved as in the magnetite case, special attention must be put on the experimental basis set. In practice a large number of reflections must be measured, precise correction of the absorption must be performed and to compare experiment and simulation, objective criterion, or confidence factor are employed. We used different criteria but focus on a criteria often used in low energy electron diffraction (LEED) which has the advantage to be a metric distance. Moreover as in LEED, one has to compare a large basis of spectra with the same type of energy range, general shape and global agreement. Thus this tool is *a priori* especially well adapted to RXS. This metric is given by

$$D_1^{(i)} = 50 \int \left| \frac{1}{c_{th}^i} S_{th}^{(i)}(e) - \frac{1}{c_{\exp}^i} S_{\exp}^{(i)}(e) \right| de$$

where (i) is the reflection index, $e$ the energy running in the intersection of the experimental and theoretical range and $S_{th,\exp}$ the theoretical and experimental spectra. They are normalized par the factor:

$$c_{th,\exp}^{(i)} = \int_{E_{\min}}^{E_{\max}} S_{th,\exp}^{(i)}(e) de$$

Then the total metric is the weighted average of the individual metric. Note that we compare thus just the shapes of the spectra and not the global relative amplitudes. The metric $D_1$ represent the area between the normalized curves in percent.

# MAGNETIC ORDERING IN V2O3

We already gave the main results of the interpretation of the experiments performed by Paolasini and coworkers. We show that the recorded spectra and azimuth scan cannot be interpreted by an orbital ordering settlement but are perfectly explained by the antiferromagnetic ordering occurring in the monoclinic low temperature phase of the compound. We want to stress here two improvements in our data analysis from this previous work. The first one concerns the integration of the non resonant magnetic scattering. This permits to improve the agreement on the (2,2,1) reflections as shown in figure 1.

REPLACE THIS TEXT WITH FIGURE GRAPHIC. TO REMOVE THIS PLACEHOLDER, CHOOSE THE "**AIP**" MENU AND SELECT "**Delete Current Single Column Section**"

**FIGURE 1.** Spectra of the two reflections in the monoclinic phase of $V_2O_3$. Top is the (221). The magnetic non resonant scattering contributes to the signal explaining the non zero intensity off the resonance. At the resonance, this term interfere with the resonant contribution. Bottom is the (111). The contributions of the important tensors are shown. The (E1-E2) magnetic dipole (1,0), measuring the toroidal moment on the atoms is the dominant term.

More important is the analysis with the help of the spherical tensors. In previous papers we already show that E1-E2 and E2-E2 channels contribute to the signal. Here we can make the distinction between the different components of each channel. As shown in the bottom of figure 1, we can see, for instance for the (1,1,1)σσ reflection that the main contribution comes from the toroidal moment. This is thus a proof of an anti-ferromagnetic ordering of the toroidal moment in the material. Their axis on each atom can also be given. Nevertheless due to the inversion center, this peculiar current cannot bring macroscopic magneto-electric comportment.

# POLAR FERRIMAGNETISM IN GAFEO3

$GaFeO_3$ is a piezoelectric material showing at low temperature a spontaneous magnetization. It reveals then strong magnetoelectric properties. Kubota and coworkers from one side and Arima and coworkers followed then the same philosophy in order to extract the physical property governing this intriguing material. In both case they performed experiment on single crystal with precise orientation of the polarization and with two and opposite direction of an applied magnetic field. The resulting spectra been the difference between the two measurements. The first authors measured the absorption spectra, the second group measured some reflections. In both cases, by this operation the unwanted contributions are eliminated and we remain with a measurement of the interesting tensors, that is making a detection of the microscopic electronic property explaining the macroscopic comportment of the material. In this case, indeed and contrary to the $V_2O_3$, the lack of inversion center makes that the magnetoelectricity is possible.

We have attempted the simulation of both experiments. We can confirm in this way the origin of the measurement (Fig. 2). Note that the diffraction experiment is more tricky because of the interference between the different component participating to the scattering. Indeed one has to consider the Thomson term of all the species presents and the interference between the different channels of the resonant process. The signal depends also on the occupancy rate, giving thus by comparison with experiment an independent way to measure it.

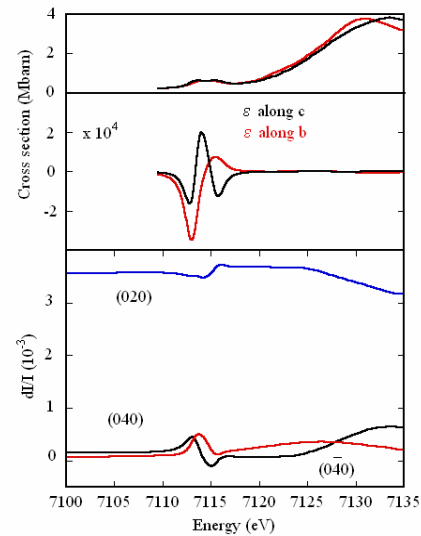

**FIGURE 2.** Simulation of the experiments performed by Kubota *et al* and Arima *et al*. On top is shown the linear dichroism effect obtained when subtracting spectra with inverted magnetic field. Bottom is shown the same effect but subtracting different reflections. Both make evidence of the toroidal moment.

## CHARGE ORDERING IN MAGNETITE

First suggested by Verwey[3] and expected in many oxides[4,5], the low temperature phase transition in magnetite has been associated to a charge disproportion on the metal atoms sites although a direct confirmation has never been evidenced in magnetite up to now. Theoretical predictions[6,7] have recently supported Verwey's scheme of the localization of the charge. However direct confirmation of charge disproportion is still lacking.

We take this example to illustrate the ability of RXS to solve quantitatively an electronic parameter. It is also a way to complete the ability of RXS were the tensor approach developed in the first section is a way to investigate these properties. It is stated there that the monopole in the E1-E1 and E2-E2 channels are a direct probe of the charge of the atoms. This is true. Nevertheless, there is another factor more important and thus more sensitive to what we want to determine. This factor is the chemical shift of the core level occurring when an atom looses or wins an electron or a fraction of electron. This results in a shift of the total structure factor. Here we want to detect a charge disproportion. One thus has advantage in choosing reflections such that the Bragg terms make that iron atoms of opposite charge subtract. Sensitivity on these so-called forbidden reflections is thus strongly enhanced. Note that this effect must be disentangled from the Tempelton effect where the anisotropy of equivalent atoms can give a quasi-equivalent effect. Thus a large number of spectra were recorded and simulations with the supposed best structure but with and without charge ordering were performed.

The most important result is that an explicit charge ordering is absolutely necessary to obtain a satisfactory agreement with the experiment (fig. 3). Reflections such as *(-4,4,1)* are highly sensitive to the magnitude of $\delta_{12}$. All of them exhibit a double structure around 7126-7131 eV which can be taken as a signature of the charge ordering phenomenon as seen by RXD. They are a clear example of the "derivative line shape", characteristic of charge ordering as likewise observed in other compounds[12,13]. Indeed, for a specific charge pattern and the reflections under consideration, the partial structure factor associated with oppositely charged iron atoms almost cancel. The optimization of the parameters at the $Fe_1$-$Fe_2$ sites gives a $\delta_{12} = 0.12\pm0.025$ electrons which corresponds to a shift of the absorption edge by 0.9 eV. This result is particularly robust and remains unchanged whether a large or a small data basis is used in the refinement. The effect of $\delta_{34}$ is contained in half-integer reflections and even more in reflections indexed in the *Cc* cell. Interestingly, and despite of the lack of knowledge about the actual lattice distortion yielding a *Cc* unit cell, the charge can be refined in this space group (see figure 1, right, for the charge distribution compatible with the *Cc* symmetry) by using reflections indexed for the *Pmca* symmetry. The value of $\delta_{34}$ is associated with a large uncertainty: $0.10\pm 0.06$ electrons, most probably reflecting the fact that each of the $Fe_3$ and $Fe_4$ positions gives rise to four inequivalent sites, being not considered in this work. To obtain a precise estimate of $\delta_{34}$ would require a concomitant knowledge of the atomic displacements producing the *Cc* structure and the use of reflections indexed in the $\sqrt{2}a_c*\sqrt{2}a_c*2a_c$ unit cell.

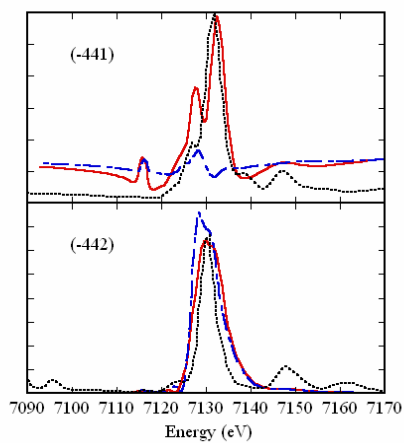

**FIGURE 3.** Two experimental and calculated resonant X-ray diffraction peak spectra in $Fe_3O_4$ at T=50K. Experimental data have been corrected for absorption. Shown are the experimental data (black dots), the calculated spectra with (red line) and without (blue line) charge ordering. Some reflections such as the (-441) clearly display the derivative effect and are therefore very sensitive to the charge ordering. Some others, as the (-442), exhibit a line shape that does not arise from the charge ordering.

## CONCLUSION

RXS is now a valuable tool to solve electronic parameter in material. Simulations are necessary to disentangle correlated parameter breaking together some symmetry. In oxides, as in most material as metallo-protein, a lot can be brought by a careful use of this technique. The resolution of the long standing problem, the charge ordering in magnetite, showing a fractionnary but important effective charge disproportion is, we hope, the path toward a generalized use of the technique.